\documentstyle[psfig]{mn}
\newif\ifAMStwofonts

\def\simgt{\mathrel{\spose{\lower 3pt\hbox{$\sim$}}
        \raise 2.0pt\hbox{$>$}}}
\def\simlt{\mathrel{\spose{\lower 3pt\hbox{$\sim$}}
        \raise 2.0pt\hbox{$<$}}}

\ifoldfss
  \ifCUPmtlplainloaded \else
    \NewTextAlphabet{textbfit} {cmbxti10} {}
    \NewTextAlphabet{textbfss} {cmssbx10} {}
    \NewMathAlphabet{mathbfit} {cmbxti10} {} 
    \NewMathAlphabet{mathbfss} {cmssbx10} {} 
  \fi
  \ifAMStwofonts
    \ifCUPmtlplainloaded \else
      \NewSymbolFont{upmath} {eurm10}
      \NewSymbolFont{AMSa} {msam10}
      \NewMathSymbol{\upi}     {0}{upmath}{19}
      \NewMathSymbol{\umu}     {0}{upmath}{16}
      \NewMathSymbol{\upartial}{0}{upmath}{40}
      \NewMathSymbol{\leqslant}{3}{AMSa}{36}
      \NewMathSymbol{\geqslant}{3}{AMSa}{3E}

      \let\geq=\geqslant \let\ge=\geqslant
    \fi
  \fi
\fi 

\ifnfssone
  \newmathalphabet{\mathit}
  \addtoversion{normal}{\mathit}{cmr}{m}{it}
  \addtoversion{bold}{\mathit}{cmr}{bx}{it}
  \newmathalphabet{\mathbfit} 
  \addtoversion{normal}{\mathbfit}{cmr}{bx}{it}
  \addtoversion{bold}{\mathbfit}{cmr}{bx}{it}
  \newmathalphabet{\mathbfss} 
  \addtoversion{normal}{\mathbfss}{cmss}{bx}{n}
  \addtoversion{bold}{\mathbfss}{cmss}{bx}{n}
  \ifAMStwofonts
    \ifCUPmtlplainloaded \else
      \UseAMStwoboldmath
      \makeatletter
      \new@mathgroup\upmath@group
      \define@mathgroup\mv@normal\upmath@group{eur}{m}{n}
      \define@mathgroup\mv@bold\upmath@group{eur}{b}{n}
      \edef\UPM{\hexnumber\upmath@group}
      \new@mathgroup\amsa@group
      \define@mathgroup\mv@normal\amsa@group{msa}{m}{n}
      \define@mathgroup\mv@bold\amsa@group{msa}{m}{n}
      \edef\AMSa{\hexnumber\amsa@group}
      \makeatother
      \mathchardef\upi="0\UPM19
      \mathchardef\umu="0\UPM16
      \mathchardef\upartial="0\UPM40
      \mathchardef\leqslant="3\AMSa36
      \mathchardef\geqslant="3\AMSa3E

      \let\geq=\geqslant \let\ge=\geqslant
    \fi
  \fi
\fi 

\ifnfsstwo
  \DeclareMathAlphabet{\mathbfit}{OT1}{cmr}{bx}{it}
  \SetMathAlphabet\mathbfit{bold}{OT1}{cmr}{bx}{it}
  \DeclareMathAlphabet{\mathbfss}{OT1}{cmss}{bx}{n}
  \SetMathAlphabet\mathbfss{bold}{OT1}{cmss}{bx}{n}
  \ifAMStwofonts
    \ifCUPmtlplainloaded \else
      \DeclareSymbolFont{UPM}{U}{eur}{m}{n}
      \SetSymbolFont{UPM}{bold}{U}{eur}{b}{n}
      \DeclareSymbolFont{AMSa}{U}{msa}{m}{n}
      \DeclareMathSymbol{\upi}{0}{UPM}{"19}
      \DeclareMathSymbol{\umu}{0}{UPM}{"16}
      \DeclareMathSymbol{\upartial}{0}{UPM}{"40}
      \DeclareMathSymbol{\leqslant}{3}{AMSa}{"36}
      \DeclareMathSymbol{\geqslant}{3}{AMSa}{"3E}

      \let\geq=\geqslant \let\ge=\geqslant
    \fi
  \fi
\fi 

\ifCUPmtlplainloaded \else
  \ifAMStwofonts \else 
    \def\upi{\pi}
    \def\umu{\mu}
    \def\upartial{\partial}
  \fi
\fi

\title[The shallow slope of the $z\sim6$ quasar luminosity function]{The shallow slope of the $z\sim6$ quasar luminosity function: limits from the lack of 
multiple image gravitational lenses}
\author[J. S. B. Wyithe et al.]
  {J.~S.~B.~Wyithe \\
  School of Physics, The University of Melbourne, Parkville, Vic, 3052, 
Australia\\
 Email: swyithe@isis.ph.unimelb.edu.au}
\date{Accepted Received}
\pagerange{\pageref{firstpage}--\pageref{lastpage}}
\pubyear{2003}

\def\LaTeX{L\kern-.36em\raise.3ex\hbox{a}\kern-.15em
    T\kern-.1667em\lower.7ex\hbox{E}\kern-.125emX}

\begin{document}

\label{firstpage}

\maketitle

\begin{abstract}
We place a limit on the logarithmic slope of the luminous quasar luminosity 
function at $z\sim6$ of $\beta\ga-3.0$ (90\%) using gravitational lensing constraints to build on the limit
of $\beta\ga-3.3$ (90\%) derived from an analysis of the luminosity distribution (Fan et al.~2003). 
This tight constraint is obtained by noting that of the two quasars which are 
lensed by foreground galaxies, neither are multiply imaged. These observations are surprising if the 
luminosity function is steep because magnification bias results in an overabundance of multiply 
imaged relative to singly imaged lensed quasars. Our Bayesian analysis uses the a~posteriori information 
regarding alignments with foreground galaxies of the two lensed quasars, and provides a constraint on 
$\beta$ that is nearly independent of the uncertain evolution in the lens population. The results suggest 
that the bright end of the quasar luminosity function continues to flatten out to $z\sim6$, as is observed 
between $z\sim3$ and $z\sim5$ (Fan et al.~2001). Provided that SDSS J1148-5251 at $z=6.37$ is magnified by
an intervening lens galaxy at $z\sim5$ (White et al.~2003), we also show that the high lens redshift in this
system implies a co-moving density of massive galaxies that is close to constant out to high redshift.
This is in agreement with the lack of redshift evolution in the velocity function of dark-matter halos with
velocity dispersions near $200$km$\,$sec$^{-1}$ 
as predicted by the Press-Schechter formalism. The combination of constraints on the quasar luminosity 
function and lens galaxy evolution are used to compute an improved estimate for the $z\sim6$ multiple 
image lens fraction of $\sim1-3\%$.

\end{abstract}


\section{Introduction}

The quasar luminosity function (LF) is the most basic property of the quasar population. At low redshifts several decades
of study have yielded a well defined optical quasar LF with powerlaw slopes at both the faint and 
bright ends that do not evolve out to $z\sim 3$ (e.g. Boyle, Shanks \& Peterson~1988; Hartwick \& 
Schade~1990; Pei 1995a; Boyle et al.~2000). At higher redshifts only very bright quasars are currently observable
at the magnitude limit of large surveys, 
and recent evidence suggests that the slope of their LF is significantly shallower than 
observed at $z\la3$ (Schmidt, Schneider, Gunn 1995; Fan et al.~2001a). 
There are now six quasars known with redshifts $z\ga5.8$ (Fan et al.~2001b,2003). 
These very high redshift quasars provide important constraints
for studies of structure formation (Turner~1990; Haiman \& Loeb~2001) and reionisation 
(e.g. Madau, Haardt \& Rees~1999; Wyithe \& Loeb~2003). Determination of their LF 
(Fan et al.~2003) is critical if we are to address these issues.  

As a population of sources for gravitational lensing the $z\sim6$ quasars
are unique, being the only sample where the gravitational lensing probability
may be of order unity (Wyithe \& Loeb~2002a,b; Comerford, Haiman \& Schaye~2002). The high expected lensing 
rate arises through a large magnification bias which increases the fraction of gravitational lenses at a 
given flux level by drawing sources from the fainter, more numerous population into a flux limited sample.
As a result the fraction of quasars in a sample that are multiply imaged by gravitational lenses
 is sensitive to the slope of the LF. Conversely, the observed multiple image lens fraction may 
be used to limit the unknown slope of a LF. This exercise was undertaken by Fan et al.~(2003). 
They presented likelihood functions for $\beta$ given the absence of multiply imaged quasars in the $z\sim6$ 
sample, and found the lack of lenses to be surprising at the $\sim90\%$ level if $\beta\la-3.5$. 
This lensing constraint is consistent 
with their findings for $\beta$ through direct analysis of the luminosity distribution in the sample which yielded
$\beta\ga-3.35$ ($90\%$). However there is a
complication. The lens fraction is linearly related (nearly) to the efficiency of the lens population,
which is proportional to the expectation value of the velocity dispersion to the fourth power and to the
space density of galaxies. Moreover, the lensing rate requires extrapolation of the local galaxy 
population to higher redshifts, and is also sensitive to cosmology (e.g. Kochanek~1996). 
Thus the unbiased lensing cross-section is quite uncertain. 

While high resolution imaging data (Fan et al.~2001b,2003) shows none of the six $z\sim6$ 
quasars to be multiply imaged, galaxies have been detected near the line of sight to two quasars 
(Shioya et al.~2002; White et al.~2003). This is puzzling because magnification bias should result in 
highly magnified multiply imaged sources being over represented among a population 
of quasars whose images are located near foreground galaxies. The effect becomes larger as the quasar 
LF becomes steeper. In this paper we present a Bayesian analysis that employs information on a~posteriori 
alignments of the two lensed quasars. This statistic is much less sensitive to the uncertainties in the lens 
cross-section than the fraction of multiply imaged quasars, and we show that it produces a tighter 
limit on $\beta$.

The paper is set out as follows. In \S~\ref{crosssec} we compute the gravitational lens 
cross-section in light of the recently measured velocity function of galaxies, 
the probability of multiple imaging for different LFs and the limits on the quasar LF that 
result from the observed lack of multiply imaged quasars. We then discuss a Bayesian approach
to computing the lens fraction in \S~\ref{bayes}. In \S~\ref{twolens} we 
describe the two high redshift quasars thought to be magnified by gravitational lensing. 
The probability of getting a multiply imaged 
quasar within a sub-sample of quasars observed to be near a lens galaxy is discussed in
\S~\ref{imageratios}. These probabilities are used to compute likelihood functions for the 
fraction of lensed singly imaged quasars, and to derive limits on $\beta$. In \S~\ref{evol} we
use the redshifts of the lens galaxies to constrain simple parametric models for the evolution 
of the lens galaxy population. We then combine these results with the limits on $\beta$ to 
estimate the expected multiple image fraction for $z\sim6$ quasars in \S~\ref{multrate}. 
Finally in \S~\ref{multim} we discuss the implications of possible multiple imaging in 
SDSS J1148-5251 for $\beta$ before presenting our conclusions in 
\S~\ref{conclusion}. Where required we assume the most recent cosmological parameters obtained through fits 
to {\em WMAP} data (Spergel et al.~2003). These include density parameters of
$\Omega_{m}=0.27$ in matter, $\Omega_{b}=0.044$ in baryons,
$\Omega_\Lambda=0.73$ in a cosmological constant, and a Hubble constant of
$H_0=71~{\rm km\,s^{-1}\,Mpc^{-1}}$.

\section{The gravitational lens cross-section and multiple imaging rate}
\label{crosssec}

The probability that a Spherical Singular Isothermal (SIS)  galaxy will lens
a back-ground source is proportional to the 4th power of its velocity dispersion
$\sigma$. Thus what is required to compute the cross-section for gravitational
lensing is the \emph{velocity function}. Until recently
the velocity function for early type galaxies (the dominant lens population, 
Kochanek~1996) had to be computed through combination of a
galaxy LF with the Faber-Jackson~(1976) relation. However this 
procedure ignores the intrinsic scatter in the Faber-Jackson~(1976) relation
and is an unreliable method (e.g. Kochanek~1993).
A more reliable representation is now
possible using the measured velocity dispersion function of early type
galaxies. Sheth et al.~(2003) presented the measured velocity
dispersion function for early type galaxies.
They suggested an analytic fit of the form
\begin{equation}
\phi(\sigma)d\sigma =
\phi_\star\left(\frac{\sigma}
{\sigma_\star}\right)^\alpha
\frac{\exp{[-(\sigma/\sigma_\star)^\beta]}}
{\Gamma(\alpha/\beta)}\beta\frac{d\sigma}{\sigma},
\end{equation}
where $\phi_\star$ is the number density of galaxies and
$\sigma_\star$ is a characteristic velocity dispersion. Sheth et
al.~(2003) found that the parameters $\sigma_\star$, $\alpha$ and
$\beta$ are strongly correlated with one another. From Sheth we take
$\phi_\star=(2.0\pm0.1)\times10^{-3}$~($h_{70}^{-1}$Mpc)$^{-3}$, $\alpha=6.5\pm1.0$,
$\beta=(14.75/\alpha)^{0.8}$ and
$\sigma_\star=161\Gamma(\alpha/\beta)/\Gamma[(\alpha+1)/\beta]~{\rm
km~s^{-1}}$.  

Given the Einstein Radius (ER) for an SIS
\begin{equation}
\xi_0=4\pi\left(\frac{\sigma}{c}\right)^2\frac{D_{\rm d}D_{\rm ds}}{D_{\rm s}}
\end{equation}
and a constant co-moving density of galaxies, the multiply imaged 
gravitational lens cross-section for sources at $z_{\rm s}$ is
\begin{equation}
\tau(z_{\rm s})=\int_0^{z_{\rm s}}dz\int_0^\infty d\sigma (1+z)^3\phi(\sigma) 
\frac{cdt}{dz}\pi\xi_0^2.
\end{equation}
We found the mean and twice the variance for a set of $\tau$ computed assuming the parameters
$\phi_\star$ and $\alpha$ to be distributed as Gaussian within their quoted uncertainties.
This procedure yields $\tau(z_{\rm s}=6)=(2.5\pm0.25)\times10^{-3}$. The statistical 
uncertainty of 10\% is
significantly smaller than is obtained through use of the Faber-Jackson~(1976) relation
and a LF. The value of $\tau(z_{\rm s}=6)$ obtained from the velocity function is 
a factor of $\sim3$ smaller than obtained in estimates of the lens fraction in the very high
redshift quasar samples (Wyithe \& Loeb~2002a,b; Comerford, Haiman \& Schaye~2002). While the 
implied lens fraction for these samples is still expected to be an order of magnitude higher than 
in lower redshift samples, it may result in less than one lens being expected in the current sample, hence
limiting the use of the lens fraction for constraining the slope of the LF
(Wyithe \& Loeb~2002b; Comerford, Haiman \& Schaye~2002; Fan et al.~2003).

As shown above the use of a measured velocity function reduces the statistical uncertainty in $\tau$ 
to 10\%. However the remaining systematic error due to the uncertain evolution in the lens 
population (e.g. Keeton~2002) makes limits based on the lens fraction uncertain. For example, 
if the co-moving density of galaxies drops in proportion to $(1+z)^{-\gamma}$ (density evolution)
then for $\gamma=1$ 
we find $\tau(z_{\rm s}=6)=(0.9\pm0.09)\times10^{-3}$, while for $\gamma=2$ the value drops to
$\tau(z_{\rm s}=6)=(0.4\pm0.04)\times10^{-3}$. Unless otherwise specified we assume a constant 
co-moving density of galaxies ($\gamma=0$). We consider only spherical lenses in this paper.  
Previous studies (Kochanek \& Blandford~1987; Blandford \&
Kochanek~1987) have found that the introduction of ellipticities $\la 0.2$ into
nearly singular profiles has little effect on the lensing cross-section and
image magnification. However the strong magnification bias will favour a high
fraction of 4-image lenses (Rusin \& Tegmark~2001). Finally we note that the 
$i$-band dropout quasars are selected independent of morphology, and so do not select against lenses,
though the possibility that the lens galaxy itself will prevent detection should be accounted for
(Wyithe \& Loeb~2002b).

\begin{figure*}
\vspace*{40mm}
\includegraphics{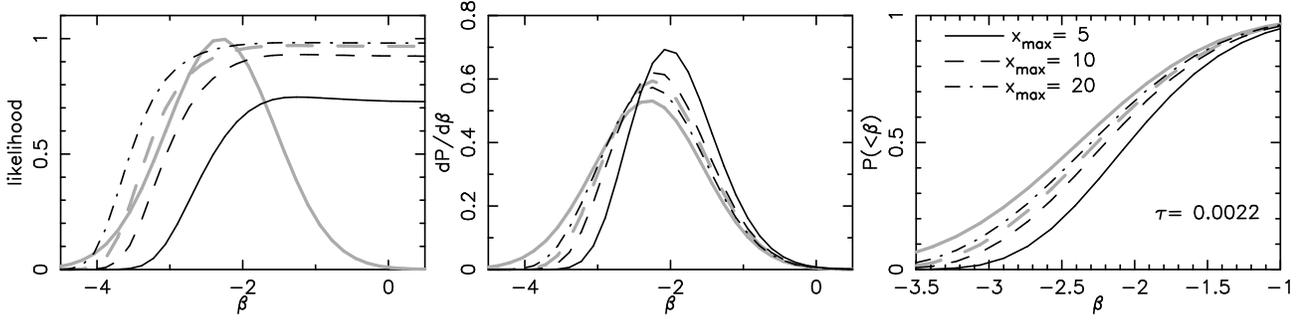}\caption{\label{fig1}Left: Likelihood functions for $\beta$. Centre: Differential probability distributions 
for $\beta$. Right: Cumulative probability distributions for $\beta$. The lensing constraints based on the 
fraction of multiply imaged quasars in the sample are shown by the dashed grey curves. 
The lensing constraints that include information on the alignments between quasars and foreground
galaxies are given for different values of $x$ (dark lines). The solid grey curves in each panel correspond 
to the likelihood and probability functions for $\beta$ based on the luminosity distribution alone (Fan et al.~2003).
Results are shown for a double powerlaw LF with $\alpha=-1.75$ and $\Delta M=4$. A constant co-moving density of
lens galaxies was assumed.}
\end{figure*}

With these points in mind we compute the a-priori probability of multiple imaging in a sample of
six quasars with $z\sim6$ for different LFs. A similar calculation has already been performed 
by Fan et al.~(2003), though we add the improvement of a more accurate $\tau$ 
computed from the velocity function, extend the calculation to account 
for constraints from the luminosity distribution (Fan et al.~2003), and provide a~posteriori limits 
on $\beta$. We also modify $\tau$ to include only
those lens galaxies which would not have contaminated the $i-$band dropout selection of the quasars
as discussed in Wyithe \& Loeb~(2002b). The probability that a quasar with luminosity $L$ and magnification
bias $B$ will be multiply imaged by a foreground galaxy is
\begin{equation}
\label{eqlf}
p_{\rm lens}(L)\sim\frac{B(L)\tau}{B(L)\tau+(1-\tau)},
\end{equation}
where $p_{\rm lens}$ is slightly overestimated because we have not included any magnification bias for 
single image quasars. 
The sum of magnifications of multiple images ($\mu$) formed by an SIS has a probability 
distribution of the form
\begin{equation}
\frac{dP_{\rm m}}{d\mu}=\frac{8}{\mu^{3}}\hspace{10mm}\mbox{for}\hspace{5mm}\mu\geq2,
\end{equation}
resulting in a magnification bias $B(L)$ for a SIS and a LF $\Phi(L)$ of
\begin{equation}
B(L)=\int_2^\infty \frac{d\mu}{\mu}\frac{8}{\mu^3}\frac{\Phi(L/\mu)}{\Phi(L)}.
\end{equation}
The probability given a LF $\Phi$, that in a sample of six quasars at $z\sim6$, 
none will be lensed is
\begin{equation} 
p_{\rm nolens}=\prod_{i=1,N_{\rm q}}\left(1-p_{\rm lens}(L_i)\right),
\end{equation}
where $N_{\rm q}=6$ and the $L_i$ are the luminosities of the $N_{\rm q}$ quasars.

\subsection{lens fractions for double powerlaw luminosity functions}
\label{lensfrac}

\begin{table*}
\begin{center}
\caption{\label{tab1}The probability given $\beta$, that in a sample of six $z\sim6$ quasars none will be lensed.
The values are tabulated assuming a constant co-moving density of lens galaxies and various values of $\alpha$ and $\Delta M$.}
\begin{tabular}{c|cccccccccccc}
\hline
$\alpha$      &  \multicolumn{4}{c}{-1.0}  &  \multicolumn{4}{c}{-1.75}  &  \multicolumn{4}{c}{-2.5}  \\
$\Delta M$  & 1         &  2     &  3   &  4     & 1       &  2  &  3       &  4     & 1        &  2  &  3       &  4    \\\hline
$\beta=-4.0$  & 0.77    & 0.42   & 0.11 &0.01    & 0.70   &  0.35  & 0.08 &0.006    & 0.47   &0.18    & 0.03 &0.002    \\
$\beta=-3.75$  & 0.82    &  0.57  & 0.28 &  0.08  & 0.77    &  0.51  & 0.24 &  0.07  & 0.56    &  0.33  & 0.13 &  0.03 \\
$\beta=-3.5$  & 0.87    &  0.70  & 0.50 &  0.32  & 0.83    &  0.67  & 0.48 &  0.29  & 0.66    &  0.51  & 0.37 &  0.23 \\
$\beta=-3.0$  & 0.92    &  0.86  & 0.78 &  0.72  & 0.89    &  0.84  & 0.78 &  0.72  & 0.78    &  0.75  & 0.72 &  0.70 \\
$\beta=-2.5$  & 0.96    &  0.93  & 0.91 &  0.90  & 0.94    &  0.93  & 0.91 &  0.90  & 0.87    &  0.87  & 0.87 &  0.87\\
$\beta=-2.0$  & 0.97    &  0.96  & 0.96 &  0.96  & 0.96    &  0.96  & 0.96 &  0.96  & 0.92    &  0.91  & 0.90 &  0.89\\\hline
\end{tabular}
\end{center}
\end{table*}

A successful fit to the low redshift ($z\la2$) quasar LF is the double powerlaw
(e.g. Boyle et al.~2000)
\begin{equation}
\label{doublepower}
\Phi(L)=\frac{\Phi_0}{L^{-\alpha}+L^{-\beta}},
\end{equation}
where $\alpha$ and $\beta$ (note we have defined $\alpha$ and $\beta$ to be negative) 
are the slopes of the faint and bright ends of the quasar LF 
respectively, and we have expressed the luminosity $L$ in units of the characteristic break luminosity. 
Use of the double power-law form for $\Phi(L)$ implies that we must specify two additional 
parameters before deriving limits on $\beta$. Firstly, since high magnifications will draw quasars that 
are fainter than the break into the sample, the magnification bias will depend on $\alpha$. 
Moreover use of the double powerlaw requires specification of the quasar luminosity with respect to the 
break. The $z\sim6$ quasars range in luminosity from $M_{1450}=-27.15\rightarrow-27.90$ (Fan et al.~2001b,2003). 
In this paper we specify the luminosity of the quasars relative to the LF in terms of the difference
between the magnitude of the faintest quasar and that of the break ($\Delta M=2.5\log_{10}L$). 
For example if the break were at $M_{1450}=-26.15$ then $\Delta M=1$.

The resulting probability $p_{\rm nolens}$ that in the sample of six $z\sim6$ quasars, none will be 
multiply imaged (given $\alpha=-1.75$, $\Delta M=4$) is plotted as a function
of  $\beta$ in the left hand panels of 
figure~\ref{fig1} (thick dashed grey line). The lack of multiply imaged quasars is only surprising at 
the 50\% level for values of $\beta<-3.3$. 
The probability $p_{\rm nolens}$ is also tabulated in Table~\ref{tab1} for various values of $\alpha$, 
$\beta$ and $\Delta M$. The lack of multiply imaged quasars is only surprising at the 10\% level if 
$\beta\la-4$ and $\Delta M\ga3$ or if $\beta\la-3.75$ and $\Delta M\ga4$.
We note that these probabilities (and those in figure~9 of Fan et al.~2003) represent a likelihood 
function for $\beta$ rather than direct to limits on $\beta$. We now turn to computation of these limits.

The posterior probability for $\beta$ is
\begin{equation}
\left.\frac{dP}{d\beta}\right|_{\rm N_{\rm mult}=0}=N p_{\rm nolens}\frac{dP_{\rm prior}}{d\beta},
\end{equation} 
where $\frac{dP_{\rm prior}}{d\beta}$ is the prior probability for $\beta$, and N is a normalising constant. 
We assume that the prior probability for the slope is flat between two bounds $\beta_{\rm min}$ and
$\beta_{\rm max}$, hence
\begin{equation}
\frac{dP_{\rm prior}}{d\beta}=(\beta_{max}-\beta_{\rm min})^{-1}.
\end{equation}  
The absence of multiply imaged quasars in the sample is surprising if $\beta$ is small because
the multiple image magnification bias is large in that case, but is less surprising as $\beta$ is increased.
Hence the likelihood function $p_{\rm nolens}$ is increasing with $\beta$ so that the fraction of 
multiply imaged quasars carries no information on its upper limit. 
As a result, the confidence with which a small value of $\beta$ can be excluded given the lensing observations 
alone is sensitive to $\beta_{\rm max}$. This dependence implies that a small probability $p$ given some $\beta$ 
that a sample of quasars would contain no lenses {\em does not} 
translate to a limit on $\beta$ at the $100(1-p)\%$ level. A second, independent constraint that limits the upper
bound on $\beta$ and hence lowers the dependence on $\frac{dP_{\rm prior}}{d\beta}$ is required.

\subsection{the addition of constraints from the luminosity distribution}
\label{limit}

Fan et al.~(2003) have derived the constraint $\beta\pm\Delta\beta$ where $\beta=-2.3$ 
and $\Delta\beta=0.8$ from their analysis of the luminosity distribution of the $z\sim6$ quasars. 
We take this bound to be the 1-$\sigma$ level of a Gaussian 
likelihood function for $\beta$. We have plotted this likelihood function for $\beta$ (normalised
to a maximum of 1), as well as the differential and cumulative probability distributions for $\beta$ in the in the 
left-hand, central and right-hand panels of figure~\ref{fig1} (solid grey lines).

We now have two constraints on $\beta$, one from the fraction of multiply imaged quasars and one
from the distribution of luminosities. Assuming these constraints to be independent (this 
assumption is discussed below), we find a joint likelihood function, and hence posterior probability 
distribution for $\beta$
\begin{equation}
\left.\frac{dP}{d\beta}\right|_{\rm N_{\rm mult}=0}=N p_{\rm nolens}\exp\left(-\frac{1}{2}\left(\frac{\beta-\bar{\beta}}{\Delta\beta}\right)^2\right)\frac{dP_{\rm prior}}{d\beta}.
\end{equation} 
The likelihood function is now normalisable because the distribution of luminosities constrains large 
values of $\beta$ (Fan et al.~2003), and is quite insensitive to the prior 
$\left.\frac{dP}{d\beta}\right|_{\rm N_{\rm mult}=0}$ as a result. We plot the 
posterior differential and cumulative probability distributions in the central and right hand panels of
figure~\ref{fig1}. These are shown by the dashed light lines, and should be compared to the 
 posterior probability distributions based on the distribution of luminosities alone (solid 
light lines). The lensing constraint disfavours smaller values of $\beta$, resulting in a narrower
probability distribution for $\beta$. The most likely value for the slope is $\beta\sim-2.2$. 
The addition of the constraint from the fraction of multiply imaged lensed quasars
improves on the limits obtained by Fan et al.~(2003) from the distribution of luminosities alone.
For this choice of $\Delta M$ and $\alpha$, the lack of multiply imaged quasars rules out 
$\beta\la-3.0$ at
the 90\% level. We also construct posterior cumulative probabilities for various values of $\alpha$ and 
$\Delta M$ and plot the results in figure~\ref{fig2} (thick dashed light lines). The addition of lensing 
constraints significantly improve the LF limits provided that the quasars are not too close to the LF 
break. For a constant co-moving density of lens galaxies (upper two rows) 
we find $\beta\ga-3.2\rightarrow-3.0$ (90\%) for $\Delta M\ga2$.

It should be noted that the two constraints are not quite independent. 
In general gravitational lensing tends to flatten the slope of the quasar LF by drawing populous 
faint quasars into a bright quasar sample (e.g. Pei~1995b). However there are 
two reasons to think that this is not a problem within the very high redshift quasar sample. 
Firstly the objects that are lensed are not multiply imaged and so have magnifications smaller than $\sim2$. 
Secondly, the average change of slope is $\Delta \beta\sim0.2$ even in the most optimistic lensing 
scenario (where $\beta\sim-3.5$) for the $z\sim6$ quasars (Wyithe \& Loeb~2002b).

\section{Bayesian approach to computing the multiple image fraction}
\label{bayes}

We may also use a Bayesian approach to compute the multiple image fraction. This will provide us with a natural
framework within which to add additional a~posteriori information on alignments of quasars with foreground galaxies
in \S~\ref{imageratios}.
Consider sources with unlensed impact parameters (in units of the ER) $y=x-1$ with associated 
magnifications $\mu$. We write the likelihood per logarithm of $x$ of observing a singly imaged lensed quasar 
(including magnification bias)
\begin{equation}
L_{\rm single}=x(x-1)\frac{1}{\mu_{\rm single}}\frac{\Phi(L/\mu_{\rm single})}{\Phi(L)},
\end{equation}
where the factor $(x-1)$ accounts for the additional solid angle available at large $y$, and 
$\mu_{\rm single}=x/(x-1)$. This likelihood may be compared to the corresponding average likelihood of 
observing a multiply imaged quasar
\begin{eqnarray}
\noindent
L_{\rm mult}&=&\int_{0}^1dy\,y\frac{1}{\mu}\frac{\Phi(L/\mu)}{\Phi(L)}\\
&=&\frac{1}{2}\int_2^\infty \frac{d\mu}{\mu}\frac{8}{\mu^3}\frac{\Phi(L/\mu)}{\Phi(L)}=\frac{B(L)}{2}.
\end{eqnarray}
The likelihood that a quasar will be singly imaged at $x$ rather than multiply imaged is therefore
\begin{equation}
\label{psingle}
p_{\rm single}(\beta\left.\right|x)=\frac{L_{\rm single}}{L_{\rm single}+L_{\rm mult}}.
\end{equation}
We may also calculate the posterior probability that a quasar will not be lensed   
\begin{equation}
p_{\rm nolens,bayes}=\int_2^\infty dx p_{\rm single}(\beta|x)\frac{dP_{\rm prior}}{dx},
\end{equation}
where $\frac{dP_{\rm prior}}{dx}$ is the prior probability for the $x$. This quantity is one minus 
the lens fraction and may be approximated using the usual formula for the lens fraction $\tau B$ or more 
accurate forms such as equation~(\ref{eqlf}). 

For the sample of $z\sim6$ quasars, the posterior probability distribution for $\beta$ is
therefore
\begin{eqnarray}
\label{problimit2}
\nonumber
\frac{dP}{d\beta}&=&N\exp\left(-\frac{1}{2}\left(\frac{\beta-\bar{\beta}}{\Delta\beta}\right)^2\right)
\frac{dP_{\rm prior}}{d\beta}\\
&\times&  \left[\prod_{i=1,6}\left(\int_2^\infty dx\,p_{\rm single}(\left.\beta\right|x,L_i)\frac{dP_{\rm prior}}{dx}\right)  \right],
\end{eqnarray}
where the prior probability distribution for $x$ can be computed from the derivative of the 
Poisson probability that a source lies within a circle of radius $x-1$ around a randomly positioned galaxy
\begin{equation}
\label{xprior}
\frac{dP_{\rm prior}}{dx}=2\tau(x-1)e^{-\tau(x-1)^2}.
\end{equation}
Equation~(\ref{problimit2}) yields identical limits to those based on the multiple image fraction (dashed light
lines in figures~\ref{fig1} and \ref{fig2}) as computed in the usual way from equation~(\ref{eqlf}). Moreover,
the magnification distribution for singly imaged sources is naturally normalised within the formalism, 
and hence the magnification bias of singly imaged sources is directly included in the calculation.

\section{Two lensed $z\sim6$ quasars}
\label{twolens}

While none of the six $z\sim6$ quasars discovered by Fan et al.~(2001b,2003) in the Sloan Digital Sky 
Survey data are multiply imaged, two have close alignment with a foreground
 galaxy, implying that they are moderately magnified. We therefore refer to these quasars
as {\em lensed}, though neither is multiply imaged.

\begin{itemize}

\item {\bf SDSS J1044-0125 at $z=5.74$:} Shioya et al.~(2003) have reported a faint foreground galaxy 
with $z\sim1.5-2.5$ at a separation of $\theta=1.9''$. Shioya et al.~(2003) estimate the velocity 
dispersion to be $\sigma\sim140-280$km/sec for this redshift range. A second image would be detectable
if $\sigma\ga220$km/sec. For a SIS galaxy this implies that the 
magnification of the image could be as high as $\mu=2$, and that the image is located at $x\sim2-10$ 
ER (where $x\sim10$ER corresponds to a $\sigma\sim140$km/sec SIS at $z\sim2.5$).

\item {\bf SDSS J1148-5251 at $z=6.37$:} This is one of two quasars known to exhibit Gunn-Peterson 
absorption troughs. White, Becker, Fan \& Strauss~(2003) present spectra showing emission features
in the Ly~$\beta$ trough which they interpret as being Ly~$\alpha$ emission from a foreground galaxy
at $z\sim4.9$. This is a likely scenario if gravitational lensing is important and was predicted by 
Wyithe \& Loeb~(2002b). Indeed, the smaller than expected Stromgren sphere implies that this quasar 
has been magnified by the intervening galaxy (White et al.~2003). 
While the alignment is presumably high, we do not know the degree of alignment with the foreground
galaxy, and hence we cannot know $x$, though is the quasar is not multiply imaged it must be larger 
than 2, and is probably comparable to 
the case SDSS J1044-0125. White et al.~(2003) do however provide an estimate of $\sigma\sim250$km/sec 
for the foreground galaxy from the velocity structure seen in the C~IV absorption system. 
For an SIS at $z=4.9$ lensing a source at $z=6.37$ with a separation $\theta$ we have 
\begin{equation}
x=4\left(\frac{\theta}{1''}\right)\left(\frac{\sigma}{250\mbox{km/sec}}\right)^{-2}\mbox{ER}.
\end{equation}
\end{itemize}

\section{improved limits on $\beta$ from close alignments with foreground galaxies}
\label{imageratios}

\begin{figure*}
\vspace*{170mm}
\includegraphics{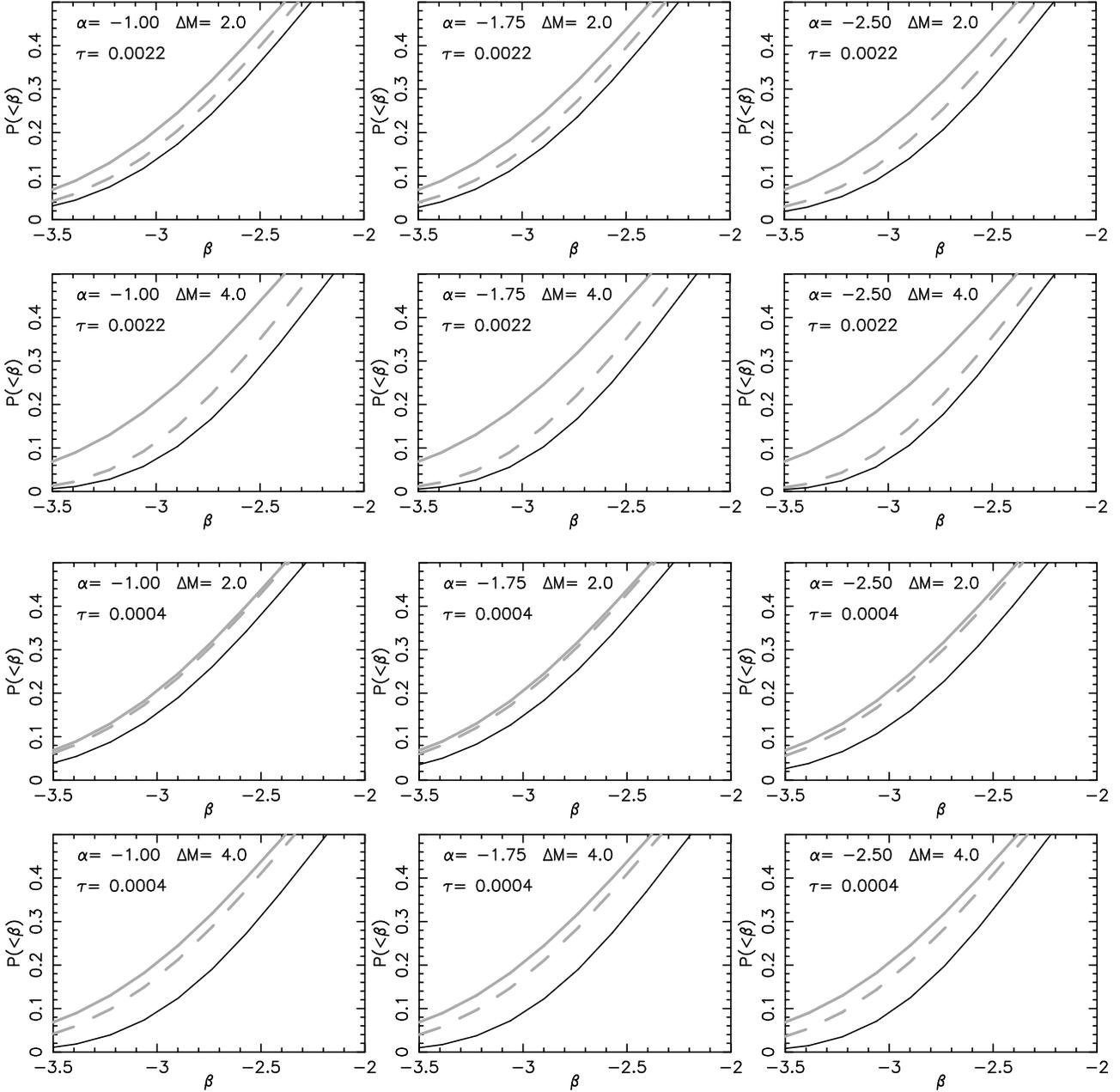}\caption{\label{fig2}Cumulative probability distributions for $\beta$. The lensing constraints based on the 
fraction of multiply imaged quasars in the sample are shown by the dashed grey curves. 
The lensing constraints that include information on the alignments between quasars and foreground
galaxies are denoted by the dark lines. The solid grey curves correspond 
to the probability functions for $\beta$ based on the luminosity distribution alone (Fan et al.~2003).
Results are shown for a double powerlaw LF for various values of $\alpha$ and $\Delta M$. The upper and
lower two rows show results assuming density evolution with $\gamma=0$ and $\gamma=2$ respectively.}
\end{figure*}

The Bayesian approach (\S~\ref{bayes}) to computing limits on $\beta$ from the multiple image 
fraction allows us to include the a~posteriori information on the alignments of the two {\em lensed} 
the $z\sim6$ quasars. 

For illustration, we begin with a hypothetical sample of 6 quasars with $\Delta M=4$, $\alpha=-1.75$ 
and (lensed) impact parameter $x$. The relative likelihoods for different values of $\beta$ given the 
lack of multiple images are $\left[p_{\rm single}(\beta\left.\right|x)\right]^{N_{\rm q}}$ as specified in
equation~(\ref{psingle}) where $N_{\rm q}=6$. These likelihoods are shown in the left hand panel of 
figure~\ref{fig1} for values of $x$ ranging from 5 to 20 (dark lines). Smaller values of $\beta$
are strongly disfavoured, particularly if $x$ is not too large. 
Next we find find the joint likelihood function and hence a posterior probability distribution for 
$\beta$ given a common impact parameter $x$ for six quasars that are not multiply imaged 
\begin{equation}
\left.\frac{dP}{d\beta}\right|_{x}=N[p_{\rm single}]^{N_{\rm q}}\exp\left(-\frac{1}{2}\left(\frac{\beta-\bar{\beta}}{\Delta\beta}\right)^2\right)\frac{dP_{\rm prior}}{d\beta}.
\end{equation} 
The resulting posterior differential and cumulative probability distributions are shown in the central and right
hand panels of figure~\ref{fig1} (dark lines). The most likely value 
is near $\beta\sim-2$ and the cumulative distributions suggest that 
$\beta\ga-3.1\rightarrow-2.7$ at the 90\% level where the systematic dependence is on $x$. 
Thus any additional information about close alignments produces constraints that may significantly 
tighten the lower limits on the slope of the $z\sim6$ quasar LF, with improvements
in the limit that are greater than 0.5 units in $\beta$ for cases where the alignment is high. 

\begin{figure*}
\vspace*{110mm}
\includegraphics{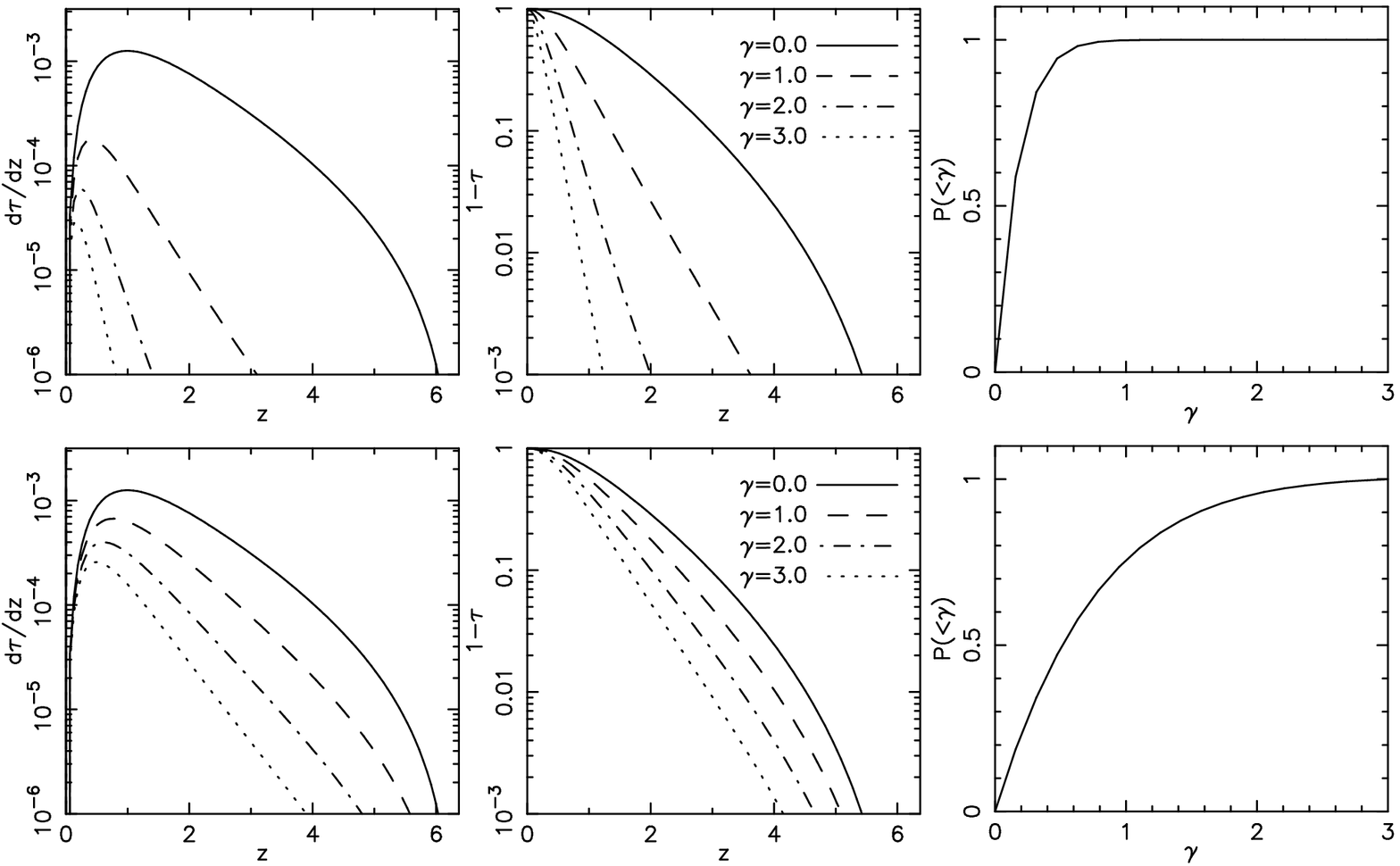}\caption{\label{fig3}Left: The differential lens cross-section for a source at $z=6.37$ for different values of $\gamma$. Centre: The corresponding fraction of lens galaxies at redshifts larger than $z$. Right: The posterior cumulative probability for $\gamma$. The upper and lower rows correspond to velocity and density evolution respectively. }
\end{figure*}

The above example suggests strong dependence of the limits derived for $\beta$ on the value of $x$. For the two
{\em lensed} SDSS quasars discussed in \S~\ref{twolens}, there are observational limits on $x$ in
the form
\begin{equation}
\frac{dP}{dx} = L_x\frac{dP_{\rm prior}}{dx},
\end{equation}
where $L_x$ is the likelihood for $x$ given the observations of $\sigma$ and $z$ for the lens galaxy, and
$\frac{dP_{\rm prior}}{dx}$ is the prior probability for $x$. Given the relation $x=x(\sigma,z)$ the
likelihood $L_x$ is
\begin{equation}
L_x= L_\sigma L_z \frac{\partial x}{\partial \sigma} \frac{\partial x}{\partial z} \frac{dP_{\rm prior}}{d\sigma}\frac{dP_{\rm prior}}{dz}.
\end{equation}
For SDSS~J1044-0215 Shioya et al.~(2003) find the majority of the dependence in the likelihoods for $\sigma$ and $z$
to be systematic, while there is no information in this regard for SDSS~J1044-0215 (White et al.~2003). 
We assume flat distributions $\frac{dP}{dx}$ with limits of $2<x<10$ for the two lensed quasars.
For the other 4 quasars we assume the prior probability distribution for $x$ (equation~\ref{xprior}).

The posterior differential probability distribution for $\beta$ then becomes
\begin{eqnarray}
\label{problimit}
\nonumber
\frac{dP}{d\beta}&=&N\exp\left(-\frac{1}{2}\left(\frac{\beta-\bar{\beta}}{\Delta\beta}\right)^2\right)
\frac{dP_{\rm prior}}{d\beta}\\
\nonumber
&\times&  \left[\prod_{i=1,2}\left(\int_2^{10} dx_i p_{\rm single}(\left.\beta\right|x,L_i)\right)  \right]\\
&\times&  \left[\prod_{i=3,6}\left(\int_2^\infty dx\,p_{\rm single}(\left.\beta\right|x,L_i)\frac{dP_{\rm prior}}{dx}\right)  \right].
\end{eqnarray}

The resulting cumulative probability distributions are shown in figure~\ref{fig2}. The limits on $\beta$ are 
significantly tighter than those obtained from the distribution of luminosities alone except in cases where 
$\alpha$ is 
large (shallow faint end slope) {\em and} $\Delta M$ is small (so that magnifications associated with multiple images 
tend to draw quasars with $\Delta M<0$ into the sample). In addition the limits are tighter than those obtained 
through consideration of the luminosity distribution and multiple image fraction (dashed light lines). 
For a constant co-moving density of lens galaxies and $\Delta M\ga2$ we find $\beta\ga-3.1\rightarrow-3.0$, while for 
 $\Delta M\ga4$ we obtain $\beta\ga-3.0\rightarrow-2.9$ (both with 90\% confidence). Thus the tightest limits 
come from the inclusion of the a~posteriori information that two of the quasars have close 
alignment with foreground galaxies. 

A second important point regarding the limit on $\beta$ provided by equation~(\ref{problimit}) is that unlike 
the limit from the multiple image fraction, it is nearly independent of the value of $\tau$. To demonstrate
this independence we have computed constraints on $\beta$ (shown in the lower two panels of figure~\ref{fig2}) 
that assume a dependence in the co-moving density of galaxies of $(1+z)^{-\gamma}$ where $\gamma=2$
(resulting in $\tau=0.0004$).
These limits may be compared with results that assume a constant co-moving density of lens galaxies 
(in the upper two panels of figure~\ref{fig2}). The limits obtained from the multiple image fraction are 
much weaker if $\gamma=2$. On the other hand the limits that use a~posteriori observations of the quasar-lens galaxy 
alignment are quite insensitive to $\gamma$.
The reason is that the role of $\tau$ is replaced by $\frac{dP}{dx}$ for the two quasars which provide the largest 
contribution to the likelihood change between large and small values of $\beta$.

\subsection{a~posteriori choice of statistic}

We have computed limits on the value of $\beta$ using two lensing based
constraints, and a~posteriori chosen the better one. This practice becomes unfair if a large number of
different constraints are available where each produces a different limit. 
In the situation described we have two different lensing constraints. However the second constraint utilises 
additional rather than different information. Thus we are justified in choosing it a~posteriori.

\section{Limits on galaxy evolution from lens galaxy redshifts}
\label{evol}

As noted by White et al.~(2003) the candidate lens galaxy in the system SDSS J1148-5251 is 
found at an improbably high redshift, which could provide an argument against the lens hypothesis 
in this system. This is quantified in figure~\ref{fig3}, where the upper curves show the differential 
cross-section for a source at $z=6.37$ lensed by a constant co-moving density of galaxies (left panel),
and the fraction of the total cross-section that 
is found at a redshift larger than $z$ (right panel). We see that the prior probability of finding a lens 
at $z\ga4.94$ among two lensed quasars is only $\sim0.01$. However the probability of finding a lens at 
high redshift is an a-posteriori statistic, i.e. we have chosen one of a possible number
of a-priori unlikely events after the observation has already been made. 
Moreover the selection of lenses within the
sample is not uniform in redshift. In particular, since the galaxy in front of SDSS J1148-5251 
was identified spectroscopically via its Ly-$\alpha$ emission line, 
it would be more easily identified at high redshift. In addition,
the cross-sections plotted in figure~\ref{fig3} refer to multiple imaging so that 
the low probability for a high redshift lens 
results from the small size of the Einstein Ring radius at high redshift. 
In contrast, the redshift distribution 
of galaxies (with velocity dispersions larger than $\sigma$)
that lie within some specified angular separation along the line of sight to a back-ground
quasar implies that there would be 1 chance in $\sim3$ of finding such 
an aligned galaxy at $z>4.96$. The role
of magnification bias in making lensing a likely scenario probably 
results in an a-posteriori probability of 
finding a high redshift fore-ground galaxy that lies somewhere between these two extremes.

The redshift distribution of gravitational lenses may be used to constrain evolution 
in the lens galaxy population (Kochanek 1992; Ofek, Rix \& Maoz~2003). While the absolute 
probability of having observed a high redshift lens is difficult to quantify, we may
more easily discuss the relative likelihoods of observing a high redshift lens as a function of 
lens galaxy population. For definiteness we consider two parameterisations for the evolution of 
the lens galaxy population: firstly evolution of the 
characteristic velocity $\sigma_\star(z)=\sigma_\star (1+z)^{-\gamma}$, and
secondly evolution of the characteristic density $\phi_\star=\phi_\star (1+z)^{-\gamma}$, which
we term velocity and density evolution respectively. The left hand panels of figure~\ref{fig3}
demonstrate the effect on the lens cross-section of varying $\gamma$. 
Values of $\gamma$ that differ from 0 (constant co-moving evolution) result in a lens population that
is truncated at high redshift. This effect may also be seen in the central panels of figure~\ref{fig3} 
where we have plotted the fraction of cross-section at redshifts larger than $z$. Values of 
$\gamma>0$ are disfavoured by the existence of a lensing galaxy at $z=4.96$ among a sample of only two 
lens galaxies.

\begin{figure}
\vspace{60mm}
\includegraphics{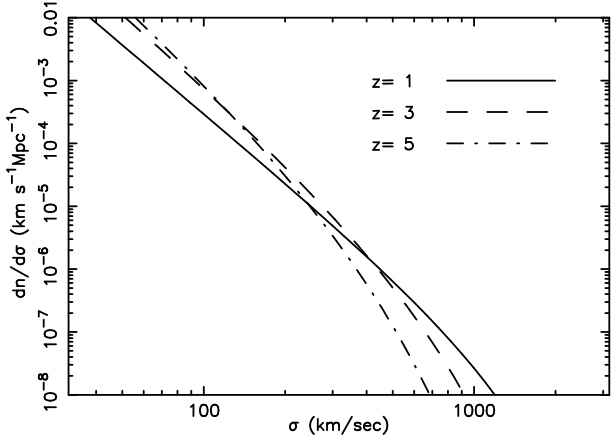}
\caption{\label{fig4}The velocity dispersion function of dark matter halos at several redshifts.}
\end{figure}

\begin{figure*}
\vspace*{170mm}
\includegraphics{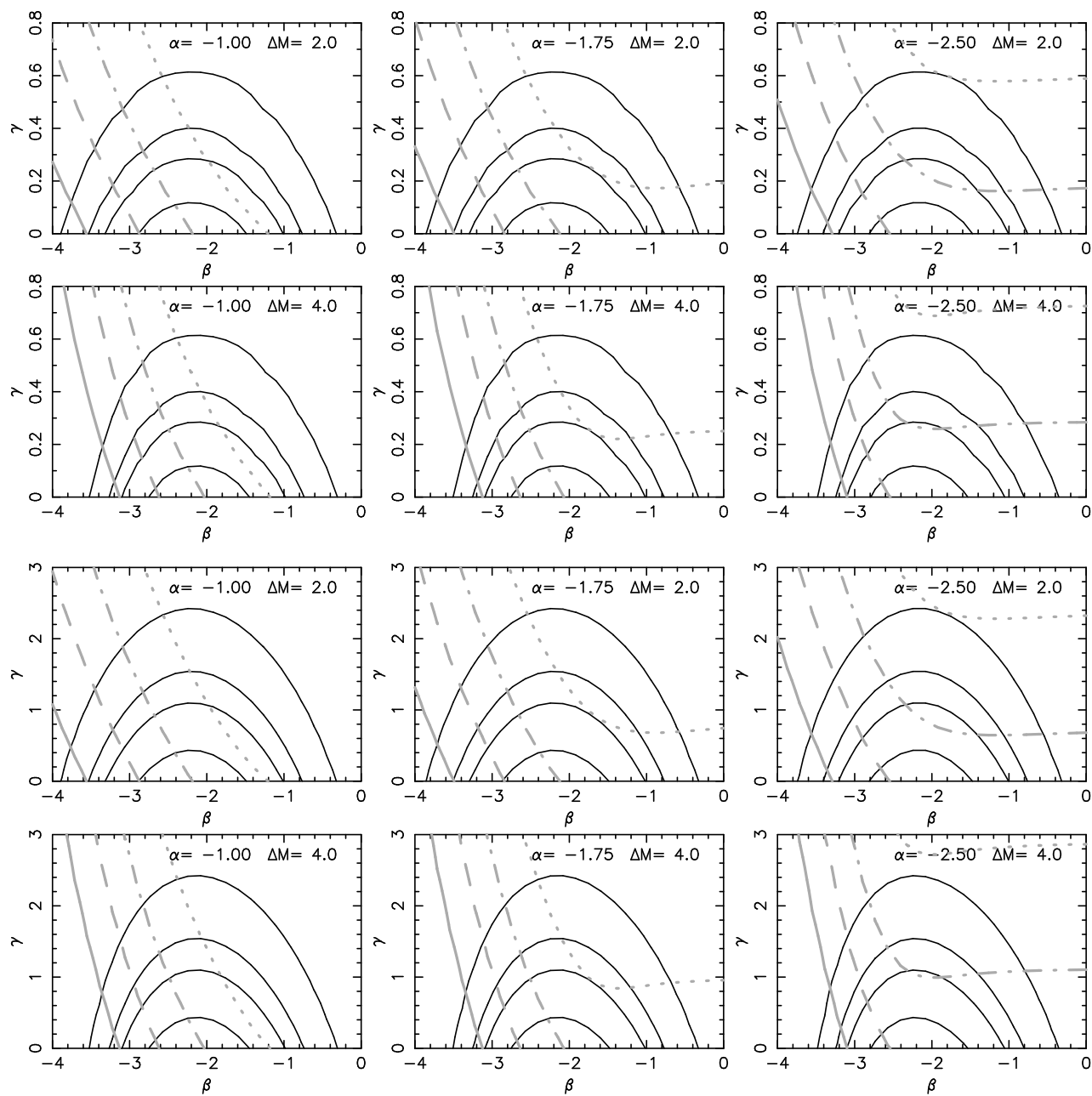}
\caption{\label{fig5} Dark contours: the joint probability function for $\beta$ and $\gamma$. The contours show
61, 26, 14 and 3.6\% of the peak value, corresponding to the 1, 2, 3 and 4 $\sigma$ levels. Grey contours: The
predicted lensing rate. The solid, dashed, dot-dashed and dotted contours correspond to lens fractions of 0.1,
0.03, 0.01 and 0.003. The upper and lower rows correspond to velocity and density evolution respectively and 
results are shown for a double powerlaw LF with various values of $\alpha$ and $\Delta M$.}
\end{figure*}

To quantify this statement we construct a likelihood function for $\gamma$ from the product of the normalised 
probabilities for the lens redshifts. The likelihood should include constants ($s_i$) to account for the relative 
detectabilities of the two lenses (the two galaxies were discovered separately via different techniques), 
though the limits on galaxy 
evolution are independent of these since the constants are independent of the evolution. The likelihood function is
\begin{equation}
\label{galLH}
L_\gamma = \prod_{i=1}^{2}s_i\int_2^\infty dx\,\frac{dP_i}{dx}(x-1)^2\frac{d\tau_i}{dz} B_i\propto \prod_{i=1}^{2}\frac{d\tau_i}{dz},
\end{equation}
where the $\frac{d\tau}{dz}$ are differential cross-sections evaluated at the 
lens and source redshifts, the integrals over the distributions $\frac{dP_i}{dx}(x-1)^2$
account for the relative alignments of the quasar and galaxy, and the $B_i$ are the magnification biases. 
The relative likelihood is dependent only on the product of the differential cross-sections. Note that
the likelihood (equation~\ref{galLH}) is only applicable if the magnification bias has aided in selection 
of the quasar, so that the source may be considered lensed. This is a caveat to the constraints imposed on
$\gamma$ in this section. However the quasar does appear to be magnified, as evidenced by the smaller than 
expected Stromgren sphere (White et al.~2003).

In the right hand panels of figure~\ref{fig3} we plot the posterior cumulative probability for $\gamma$
\begin{equation}
P(<\gamma)=\int^{\gamma}N L_{\gamma'} \frac{dP_{\rm prior}}{d\gamma'},
\end{equation}
assuming a flat prior probability distribution for $\gamma$ at values greater than $0$. By excluding
the possibility of $\gamma<0$ we are assuming that the lens galaxy population increases monotonically
in time as expected in hierarchical merging scenarios. This choice also leads to more conservative 
limits on $\gamma$. We find $\gamma\la0.4$ and 
$\gamma\la1.6$ at the 90\% level assuming velocity and density evolution 
respectively. The possible presence of a lens galaxy at such a high redshift therefore offers an opportunity
to constrain the (mass selected) co-moving density of massive galaxies to be close to
constant out to high redshifts. This result is consistent with the study of Ofek, Rix \& Maoz~(2003) who
performed a detailed study on a large sample (15) of multiple image lenses at $z\sim1-2$. 

A constant co-moving density of lens galaxies out to $z\sim5$ may 
not be surprising in light of the Press-Schechter~(1976) 
prediction for the velocity function of dark-matter halos (number per cubic comoving Mpc per unit velocity). 
Taking the circular velocity 
$v_{\rm vir}$ to equal the virial velocity of an SIS dark matter halo with mass $M$ (Barkana \& Loeb~2001), 
and a velocity dispersion $\sigma=v_{\rm vir}/\sqrt{2}$ we find
\begin{equation}
\frac{dn}{d\sigma}=M\frac{dn}{dM}\frac{\sqrt{2}}{3v_{\rm vir}},
\end{equation}  
where $\frac{dn}{dM}$ is the Press-Schechter~(1976) mass function.
The resulting velocity function of dark matter halos is plotted in figure~\ref{fig4}
at a series of redshifts. Note that near velocity dispersions 
of $\sigma\sim200$km$\,$sec$^{-1}$,
which dominate the lens cross-section, there is little evolution in $\frac{dn}{d\sigma}$ from $z\sim1$ 
to $z\sim5$. If massive galaxies occupied dark matter halos in the past as they do today, we would therefore
expect little evolution in the lens population, even out to large redshifts.

\section{What is the multiple image lensing rate for $z\sim6$ quasars?} 
\label{multrate}

\begin{figure*}
\vspace*{170mm}
\includegraphics{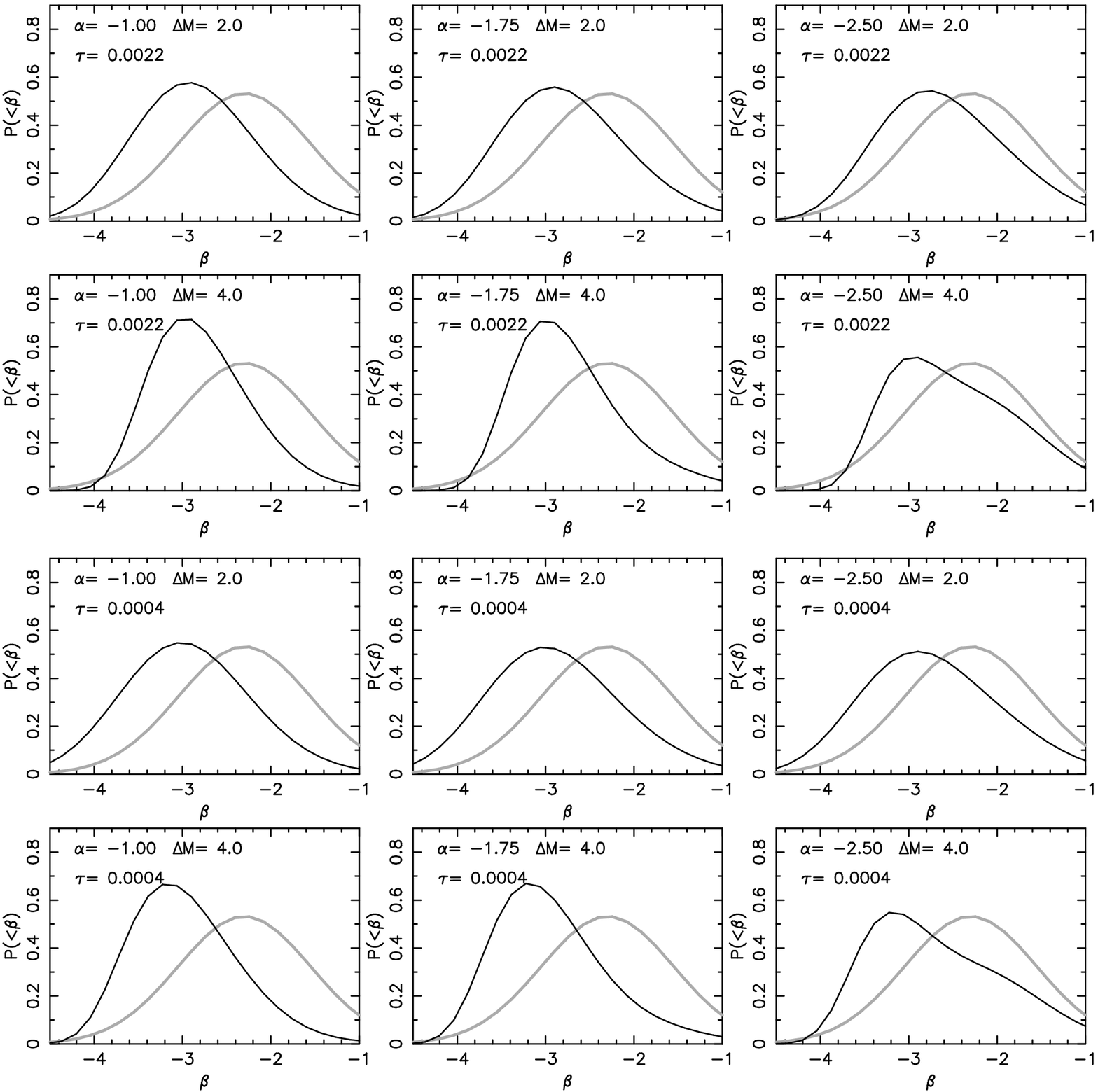}
\caption{\label{fig6} Differential probability distributions for $\beta$. 
The lensing constraints that include information on the alignments between singly imaged quasars and foreground
galaxies as well as a single multiply imaged quasar are denoted by the dark lines. 
The solid grey curves correspond 
to the probability functions for $\beta$ based on the luminosity distribution alone (Fan et al.~2003).
Results are shown for a double powerlaw LF for various values of $\alpha$ and $\Delta M$. The upper and
lower two rows show results assuming density evolution with $\gamma=0$ and $\gamma=2$ respectively.}
\end{figure*}

We may combine the information obtained for $\beta$ and $\gamma$ and estimate the expected multiple
imaging rate for $z\sim6$ quasars. Since our earlier limits on $\beta$ are independent of $\tau$, while
the limits derived for $\gamma$ are insensitive to the magnification bias, the constraints placed
on $\beta$ and $\gamma$ in \S~\ref{imageratios} and \S~\ref{evol} are independent. In figure~\ref{fig5} we have 
plotted the resulting joint probability function (dark contours); the contours shown are at 61, 26, 14 and 
3.6\% of the peak value, corresponding to the 1, 2, 3 and 4 $\sigma$ levels of a Gaussian distribution. 
The upper two and lower two rows 
of figure~\ref{fig5} correspond to velocity and density evolution respectively and in each case results are 
shown for a double powerlaw LF with various values of $\alpha$ and $\Delta M$. The figure shows that the preferred 
values are found near near $\gamma=0$ and $\beta=-2.1$. We also show contours of the multiple image
fraction (light contours); the solid, dashed, dot-dashed and dotted contours correspond to lens fractions of 0.1,
0.03, 0.01 and 0.003. We find that the 
multiple image fraction should be $\sim1-3\%$. This value is lower than previous estimates due to constraints on 
the shallow luminosity function, and unfortunately implies that a $z\sim6$ lens may not be found among the 
complete sample of SDSS $z\sim6$ quasars.

\section{What if SDSS J1148-5251 were multiply imaged?}
\label{multim}

While current high resolution imaging suggests that all of the $z\sim6$ quasars are point sources, 
White et al.~(2003) note the possibility that SDSS J1148-5251 is multiply imaged cannot be 
ruled out by current observations due to the small angular diameter of the Einstein 
ring in this system; $\Delta\theta\sim0.3''$ for a $\sigma=250$km/sec galaxy at $z=4.94$ 
lensing a quasar at $z=6.37$. For comparison the $z\sim6$ quasars have been imaged at a 
resolution of $\sim 0.4''$ (Fan et al.~2003). It is therefore possible that SDSS J1148-5251 
is multiply imaged but appears as a point source, though we note that this is an unlikely scenario 
since with $4''$ seeing, a double with a $3''$ separation should be recognisable
if the flux ratio is smaller than $10:1$ (Chris Kochanek 2003, private communication). 
Multiple imaging of SDSS J1148-5251 would have important implications for the study of the $z\sim6$ LF, 
invalidating the constraints on $\beta$ obtained in \S~\ref{imageratios}. We have therefore 
computed the limits imposed on $\beta$ by the observation of one multiply imaged source (SDSS J1148-5251) 
and one singly imaged source with 
high alignment (SDSS J1044-0125) among a sample of six quasars at $z\sim6$. 

The likelihood function for this scenario may be written as the product of the probability that a source is 
lensed with the likelihood that the remainder are singly imaged. The observation that one of the quasars is
multiply imaged constrains large values of $\beta$, while the observation of high alignment without 
multiple imaging limits small values of $\beta$ as discussed in previous sections. The maximum of the combined 
likelihood function lies in the lower end of the range specified by the luminosity distribution (Fan et al.~2003). 
The posterior probability function for the sample of $z\sim6$ quasars may be written
\begin{eqnarray}
\nonumber
\frac{dP}{d\beta}&=&N\exp{\left(-\frac{1}{2}\left(\frac{\beta-\bar{\beta}}{\Delta\beta}\right)^2\right)}\frac{dP_{\rm prior}}{d\beta}\\
&&\times p_{\rm lens}(\beta,L_1)\times \left[\int_2^{10} dx p_{\rm single}(\beta|x,L_2)\right]\\
&&\times\left[\prod_{i=3}^{6}\left(\int_2^\infty dx p_{\rm single}(\beta|x,L_i)\frac{dP_{\rm prior}}{dx}\right)\right]
\end{eqnarray}
and is plotted in figure~\ref{fig6} for various values of $\Delta M$ and $\alpha$. We find the preferred value in this 
case to be $\beta\sim-3$. As noted in \S~\ref{lensfrac} it is surprising that
one quasar in the sample would be lensed, but not at a highly significant level. The observation of one lensed
lensed quasar therefore prefers smaller values of $\beta$, for which the magnification bias
is larger, and also slightly tightens the allowed range for $\beta$. 
In summary, if SDSS J1148-5251 were multiply imaged the preferred value for the slope would be $\beta\sim-3$, 
which is ruled out at the 90\% level if the quasar is singly imaged but with a high alignment. From 
figure~\ref{fig5} we see that $\beta\sim-3$ implies a lens fraction of $\sim3-10\%$ rather than $\sim0.3-1\%$,
which is more consistent with previous estimates. This underlines the importance of determining whether this quasar is 
multiply imaged or merely lensed.

\section{conclusion}
\label{conclusion}

From their analysis of the luminosity distribution of quasars at $z\sim6$ Fan et al.~(2003) determined a slope for 
the quasar LF of $\beta\ga-3.3$ (90\%). This slope is consistent with the value found for the slope 
of the LF at $z\sim4.3$, but is not consistent with the slope of the LF of bright 
quasars at $z\la3$.
It is also possible to constrain the slope of the LF using the fraction of multiply imaged lensed 
quasars. Fan et al.~(2003) computed the probability of obtaining a lens fraction of zero 
as a function of $\beta$ and 
found that the constraints were similar to those of the LF. We have performed a Bayesian analysis 
including both of these (nearly) independent constraints, yielding the result that at 90\% confidence 
$\beta\ga-3.3\rightarrow-3.0$ provided that the quasars are at least 2 magnitudes brighter than the unknown 
position of a break in a double powerlaw LF. The systematic dependence in the constraint is due to the 
unknown slope of the LF at fainter luminosities, the luminosity of the break and the uncertain evolution in the 
lens galaxy population. 

While inclusion of constraints 
from the multiple image fraction somewhat improves the limits on $\beta$, we have shown that 
the additional information from observations that neither of the two quasars which lie near to the line of 
sight to foreground galaxies (and which are therefore lensed) are multiply imaged provides a stronger lensing 
based constraint on the slope $\beta$. We find that for a double powerlaw LF $\beta\ge-3.1\rightarrow-2.9$ 
with 90\% confidence. Unlike the constraint that uses only the multiple image fraction, this limit is 
nearly independent of evolution in the lens population, and adds further evidence of a trend to shallower LF 
slopes at large redshifts. 

We also find that the existence of a lens galaxy at $z\sim5$ in a sample of two lenses
constrains the evolution in the massive galaxy population to be close to that of constant co-moving density
(provided that the quasar behind the $z\sim5$ galaxy is subject to magnification bias). This lack of evolution 
in the lens population is consistent with the lack of redshift evolution in the velocity function of 
dark-matter halos (for velocity dispersions near 200km$\,$sec$^{-1}$) as predicted by the Press-Schechter formalism.

Finally the constraints on the quasar luminosity function and lens population have been used to compute an improved
estimate for the expected $z\sim6$ multiple image lens fraction of $\sim1-3\%$. This value is lower than previous
estimates due to the tight constraints on the slope of the LF.

\section*{Acknowledgements}
I would like to thank Chris Kochanek for helpful comments during the course of
this work.

\label{lastpage}

\end{document}